\documentclass[showpacs,amsmath,amssymb,twocolumn]{revtex4}
\usepackage{graphicx}
\usepackage{dcolumn}
\usepackage{bm}
\usepackage{amsmath}
\usepackage{amssymb}

\begin{document}

\title{Operating molecular transistors as heat pumps}

\author{D.~F.~Martinez}
\author{Bambi~Hu}
\affiliation{Department of Physics, Center for Nonlinear Studies, and The
Beijing-Hong Kong-Singapore Joint Center for Nonlinear and Complex
Systems (Hong Kong), Hong Kong Baptist University, Kowloon Tong, Hong
Kong, China, and Department of Physics, University of Houston, Houston, TX
77204-5005, USA}

\begin{abstract}
We study heat transport in transistor-like devices composed of two reservoirs and a gate electrode, with a ballistic electronic one-dimensional system connected between the two reservoirs and interacting with a laser field. We derive in a simple way an equation for the heat flux in terms of the Floquet-Green operator of the system. As a case example, we investigate a two-level transistor and find that the laser field can produce a net heat flow out of either one of the reservoirs. The direction of this flow is determined by the gate voltage.  We study numerically the dependence of this heat-pump on the relevant parameters of the system and show that there is a minimum temperature below which it does not operate.
\end{abstract}


\maketitle

\begin{figure}
\begin{center}
\includegraphics[width=7cm,height=6cm]{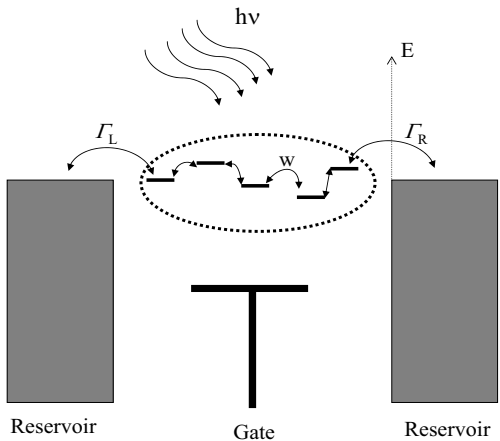}
\caption{ \label{fig1}Schematic diagram with level structure of a molecular transistor. The inter-site coupling is given by $w$ and the coupling to the reservoirs by $\Gamma_{\rm L}$ and $\Gamma_{\rm R}$. An applied gate voltage and an incident laser field allow this device to pump heat into either reservoir.}
\end{center}
\end{figure}

The fast pace of development of nano-scale technologies
and devices, has been an important driving force behind
the theoretical work on the physics at such scales.
The idea of constructing miniature versions of current
pumps, heat pumps and heat engines seems quite attractive
for its broad technological applications, from nano-machines
to molecular processors to highly efficient refrigeration
and better solar cells.
Low-dimensional systems are particularly interesting because
of their unusual properties. Given the typical
length of these systems, transport in many of these devices (e.g. carbon nanotubes) is ballistic, and can
therefore be studied within Landauer's formalism.
One important question in this context is the influence
of excitations by electromagnetic fields and gate
voltages on electron transport. Interesting phenomena
like photon-assisted tunneling and adiabatic and non-adiabatic current pumping have already been found in these kinds of systems.

Quantum heat pumping in electronic systems has been
proposed using energy filtering \cite{Humphrey}
and adiabatic pumping \cite{Moskalets, Wang}. There is also a
proposal for a molecular pump using modulation of two
vibrational levels of a molecule \cite{Segal}. More recently, in a development parallel to ours,
heat pumping in a driven two-barrier system has been reported \cite{Liliana}.

In this paper we propose that a typical transistor-like device, either molecular or quantum-dot based, composed of a
quantum system (asymmetrical) connected between two reservoirs and with a third electrode providing a gate
voltage, can be used to pump heat by simple illumination with a laser beam (THz laser for quantum dot systems,
middle infrared for molecular).
Molecular transistors have been proposed as the next step in the miniaturization of electronic devices. Our results indicate that some of these molecular-based transistors could be used to provide in-situ refrigeration in nano-processors.
In this work we show detail numerical results for the simplest system that can be used as a heat pump: a driven two-level system.  We show how this heat pumping action is strongly gate voltage dependent and bi-directional. The effect of the different parameters of the system on the performance of the device is discussed in detail.

For a non-driven device composed of two reservoirs and a connecting 1-D system, the probability that an electron on the left reservoir (L) has energy $E$ is given by the fermi distribution function $f_{\rm L} (E)$.
The probability that this electron will reach the right reservoir (R) is given by the transmission function $t(E)$, and
the amount of energy that this electron will deposit there is $E - \mu_{\rm _R}$, where $\mu_{\rm _R}$ is the chemical potential of the right reservoir. From Landauer's formula, the energy flux into R is,
\begin{equation}
Q_{\rm R,in} = 2/h \int^\infty_{-\infty} (E-\mu_{\rm _R} )t (E) f_{\rm L}(E) dE ~.
\end{equation}
Similarly, the energy flux out of R can be written as
\begin{equation}
Q_{\rm R,out} = 2/h \int^\infty_{-\infty} (E-\mu_{\rm _R} )t (E) f_{\rm R}(E) dE ~.
\end{equation}
The total energy flux out of this electrode is $Q_R =Q_{\rm R,in} - Q_{\rm R,out}$.
Similar equations are obtained for the left electrode. The resulting equations for the energy flux into the
electrodes are
\begin{equation}
\begin{aligned}
Q_{\rm R} &= ~2/h \int^\infty_{-\infty} \left( E-\mu_{\rm _R} \right) t(E)\left( f_{\rm L}(E)- f_{\rm R}(E)\right) dE \\
Q_{\rm L} &=-2/h \int^\infty_{-\infty} \left( E-\mu_{\rm _L} \right) t(E)\left( f_{\rm L}(E)- f_{\rm R}(E)\right) dE
\end{aligned}
\end{equation}

\begin{figure}
\begin{center}
\includegraphics[width=8cm,height=5cm]{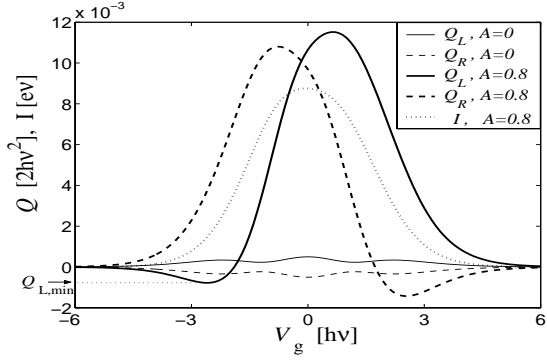}
\caption{ \label{fig2} Two-level nano-transistor illuminated with a laser field. Heat flux out of reservoirs and
pumped current vs gate voltage. Non-driven ($A=0$) and
driven case ($A=0.8$). $\Delta E = -0.8$, $T = 0.55$, $\Delta T = 0.01$. }
\end{center}
\end{figure}

In the absence of a potential difference between R and
L, $\mu_{\rm _L} = \mu_{\rm _R}$ and therefore, $Q_L = -Q_R$. Energy is transferred
completely from one electrode (the hot one) into the other (the cold one).

A time-periodic field allows particles to change their energy by any multiple of $h\nu$, where $h$ is planck's constant and $\nu$ is the frequency of the time-periodic field. The effect of a periodic driving on current transport in simple systems has been
studied for some time \cite{Gloria}. Among other interesting effects that have been found are dynamic localization, photon assisted
tunneling, adiabatic and non-adiabatic charge pumping, and others.
A treatment of the transport properties of nanoscale systems in the presence of a driving field using Floquet
theory has been presented by Kohler et al. \cite{Kohler}. A Floquet-Green operator can be defined and calculated in terms
of matrix continued fractions \cite{Martinez}. For a tight-binding 1-D system with L sites, driven by the oscillating electric field of a laser, we write the Hamiltonian (in the dipole approximation) as
\begin{equation}
\begin{aligned}
\label{Hamiltonian}
H= &\sum_{\substack {j=1}}^L \epsilon_j \left|j\right>
\left<j\right| -w\sum_{\substack {j=1}}^{L-1} \left(\left|j+1\right>\left<j\right|+
\left|j\right>\left<j+1\right|\right)+ \\
& - \rm{i}(\Gamma_{\rm L}\left|1\right>
\left<1\right| + \Gamma_{\rm R}\left|L\right>
\left<L\right|) + A\cos{2\pi\nu t}\sum_{\substack {j=1}}^L \left|j\right>j\left<j\right| .
\end{aligned}
\end{equation}
Here, the $\epsilon_j$'s represent the on-site energies, $w$ is the coupling energy between sites and $\Gamma_{\rm L(R)}$ represents the self-energy due to the coupling between site $j=1$($j=$L) and the left(right) reservoir. Also, $A=eEd$ where $e$ is the electron charge, $E$ is the amplitude of the electric field of the laser and $d$ is the characteristic inter-site distance. The frequency of the laser field is given by $\nu$.

The (time averaged) probability that a particle with energy
$E$ initially on the left electrode will move through the
system, absorb/emit $k$ photons ($k > 0$ for absorption, $k <
0$ for emission), and exit the system through the right
lead with energy $ E + k h\nu$ is
\begin{equation}
\label{eq:deflambdafloquet}
T^k _{\rm RL} (E)=\Gamma_{\rm L} \Gamma_{\rm R} \left| G^{(k,0)}_{\rm L,1}(E) \right| ^2.
\end{equation}
where $G^{k,r}_{\rm m,n} (E)$ are the components of the Floquet operator
(upper indexes specify Floquet channels, lower indexes for
spatial location). With this, we can construct, for driven systems, the equivalent of Eq.(1)-(2). The (time averaged) energy flow out of the right lead is
\begin{equation}
Q_{\rm R, out} = 2/h \int^\infty_{-\infty} (E-\mu_{\rm _R} )\sum_{\substack {k=-\infty}}^\infty T^k_{\rm LR} (E) f_{\rm R}(E) dE ~,
\end{equation}
and the energy flow into the right lead is
\begin{equation}
Q_{\rm R, in} = 2/h \int^\infty_{-\infty} (E+kh\nu-\mu_{\rm _R} )\sum_{\substack {k=-\infty}}^\infty T^k_{\rm RL} (E) f_{\rm L}(E) dE ~.
\end{equation}
From the equations above, and using $ \dot{Q}_R = \dot{ Q}_{\rm R,in} - \dot{Q}_{\rm R,out}$ we obtain the following expression for the total energy flow into the right reservoir (and similarly for the left electrode)

\begin{small}
\begin{equation}
\begin{aligned}
Q_{\rm R} &= 2/h \int_{\substack {-\infty}}^\infty {(E-\mu_{\rm _R} )\sum_{\substack {k=-\infty}}^\infty \left( T^k_{\rm RL} (E) f_{\rm L}(E)- T^k_{\rm LR} (E) f_{\rm R}(E) \right)dE} \\
&+ 2\nu \int^\infty_{-\infty}\sum_{\substack {k=-\infty}}^\infty k T^k_{\rm RL} (E) f_{\rm L}(E) dE ~~,\\
Q_{\rm L} &=-2/h \int^\infty_{-\infty} (E-\mu_{\rm _L} )\sum_{\substack {k=-\infty}}^\infty \left( T^k_{\rm RL} (E) f_{\rm L}(E)- T^k_{\rm LR} (E) f_{\rm R}(E) \right) dE \\
&+ 2\nu \int^\infty_{-\infty}\sum_{\substack {k=-\infty}}^\infty k T^k_{\rm LR} (E) f_{\rm R}(E) dE~.
\end{aligned}
\end{equation}
\end{small}
The first term on the right of both equations has the same structure as the pumped current \cite{Kohler}, but with $E-\mu_{\rm R(L)}$ replaced by the electron charge. The second term gives a warming-up effect of the reservoirs due to the presence of the time-dependent field, as was recently reported by Arrachea \textit{et. al}\cite{Liliana} using a non-linear Green-function approach.

It has been shown \cite{Kohler} that a time-periodic field can pump current in a spatially asymmetric system. Experimentally this was first demonstrated by Yasutomi \textit{et al.} \cite{Yasutomi} using dipolar molecules. The simplest theoretical system where this condition is achieved is the two-level system with different on-site energies. In such case, a laser field acting on the system can produce a current, even in the absence of a voltage difference between the reservoirs. In Eq.\ref{Hamiltonian} we take  $L=2$ and $j=-1,1$. Also, we write $\epsilon_{\pm 1}=V_{\rm g} \pm \Delta E/2$, with $V_{\rm g}$ a parameter controlled by the gate voltage, and $\Delta E=\epsilon_1 - \epsilon_{-1}$ being the on-site energy difference. We also assume equal coupling to the reservoirs, $\Gamma_{\rm L} = \Gamma_{\rm R}$.
\begin{figure}
\begin{center}
\includegraphics[width=8cm,height=5cm]{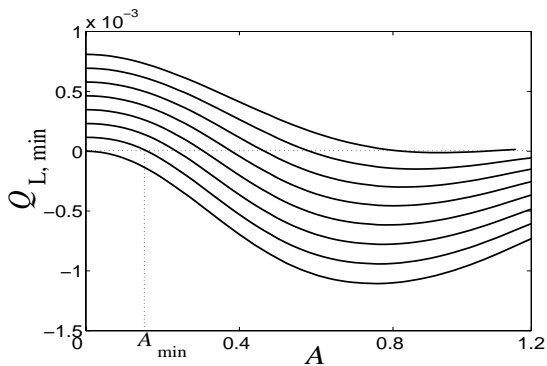}
\caption{\label{fig3}Minimum value of $Q_{\rm L,min}$ as a function of laser field
amplitude $A$, and for different $\Delta T$'s. Bottom $\Delta T = 0$, top
$\Delta T= 0.035$. $Q_{\rm L,min} = 0$ at $A = A_{\rm min}$. }
\end{center}
\end{figure}

\begin{figure}
\begin{center}
\includegraphics[width=8cm,height=5cm]{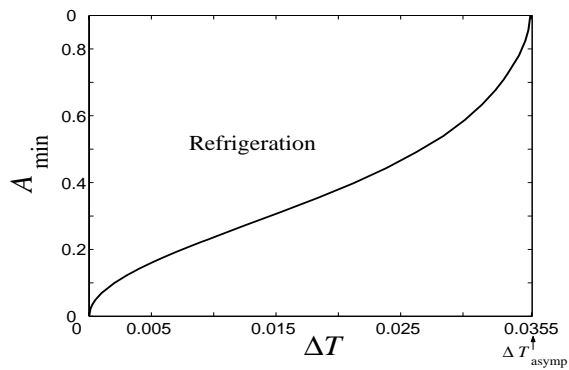}
\caption{ \label{fig4}Minimum field amplitude for the onset of refrigeration,
as a function of $\Delta T$. We use $T=0.55$. Similar curves are obtained
at other temperatures but with different $\Delta T_{asymp}$. }
\end{center}
\end{figure}
Using this Hamiltonian, we find that the pumped current
is symmetrical with respect to the gate voltage (for
$V_{\rm g} = 0$ we assume the two levels are located symmetrically
around the fermi energy of the leads) and it is antisymmetrical
with respect to the energy difference $\Delta E$.
In Fig.\ 2 we show the energy flux into the left reservoir
and the right reservoir, as well as the pumped
current, as a function of the gate voltage. All energy
parameters are given in units of $h\nu$, and temperatures
are given in units of $h\nu/k$, where $k$ is Boltzman's constant.
In this figure, $T = 0.55$, $\Delta E = -0.8$, $\Delta T = 0.01$.
In all our calculations $w = h\nu$ and $\Gamma = 0.1 w$. Also, the left
electrode is always the cold one, with $T_L = T - \Delta T /2$, and
$T_R = T + \Delta T$. The thin lines (full line for the left electrode
and dashed for the right one) show the results for the static case
$A = 0$, where $Q_R = -Q_L$, and $Q_R < 0$. For the case
$A = 0.8$ we can see a significant change in the heat fluxes,
with a minimum developing near $V_g =  -2.7$, and $Q_L <0$.
Interestingly, for this value of the gate voltage, heat is pumped out of the cold reservoir and into
the hot reservoir. In this figure there are three regimes for
heat transport depending on the gate voltage: For large
positive values of $V_{\rm g}$ heat is transported out of the hot
reservoir and into the cold reservoir. For values of $V_{\rm g}$ around $V_{\rm g} = 0$ energy from the laser field is
pumped into both reservoirs, thus warming them. Finally, for large negative values of the gate voltage, heat
is pumped out of the cold reservoir and given to the hot reservoir (in addition to the energy contributed by the
laser field). As we discuss next, this bidirectional heat pumping does not always occur. In this figure, $Q_{\rm L,min}$ is the minimum value of $Q_{\rm L} (V_{\rm g})$.

In Fig.\ 3 we show $Q_{\rm L,min}$ as a function of the amplitude $A$ of the laser
field and for different values of the temperature difference $\Delta T$. From bottom to top $\Delta T = 0, 0.005, 0.001, ...0.035$.
As expected, the cooling effect of the laser field is countered
by thermal conduction. For small values of $A$, and for $\Delta T>0$ thermal conduction dominates and therefore $Q_{\rm L,min} > 0$. Notice that the upper curve in Fig.\ 3 ends abruptly near $A = 1.1$. Beyond this point there is no longer a local minimum in $Q_{\rm L}$ as a function of $V_{\rm g}$. The amplitude $A_{\rm min}$ to reach the break-even
point ($Q_{\rm L,min} = 0$) depends on $\Delta T$, and as one would
expect, increases with it. For $A > A_{\rm min}$ there is refrigeration.

The behavior of $A_{\rm min}$ as a function
of $\Delta T$, and for an arbitrary $T = 0.55$ is shown in Fig.\ 4. A vertical asymptote
is found at $\Delta T= \Delta T_{\rm asymp}= 0.0355$ which corresponds to the
value of $\Delta T$ for which the (local) minima of $Q_{\rm L,min}$ vs
$A$ grazes the $x$ axis. There are possibly other minima at
even higher values of $A$, which would give rise to other
asymptotes in $A_{\rm min}$ $vs.$ $\Delta T$. Here we will limit ourselves
to values of $A$ up to the first asymptote. More
over, we have numerical evidence that indicates that this
asymptotic value of $\Delta T$, for $T < 0.25$, is actually the
maximum value that $\Delta T$ can have, at a given temperature, after
which no refrigeration can occur. The value of $\Delta T_{\rm asymp}$ for different temperatures is shown in Fig.\ 5. An
interesting feature in Fig.\ 5 is the existence of a minimum
value of temperature, $T_{\rm min}$, below which there was no
refrigeration at any amplitude of the laser
field. This minimum value is clearly different for different
values of the on-site energy difference $\Delta E$, a quantity
that is essential for the pumping action of this device. In
general, for most amplitudes of the laser field, greater
$\Delta E$ gives greater heat pumping power of this
device. Surprisingly, this does not translate into a monotonically
decreasing $T_{\rm min}$ with increasing (in magnitude)
$\Delta E$. As can be seen in Fig.\ 6 for $\Delta E<-0.8$, the
minimum temperature for refrigeration goes slowly up. For $\Delta E> -0.8$ this minimum temperature goes up
more rapidly, and as expected, it becomes infinite when
$\Delta E = 0$ (no pumping possible in a space symmetric, laser driven
system). The minimum temperature for operation
of this device is near $T_{\rm min}= 0.065~[h\nu/k]$.

\begin{figure}
\begin{center}
\includegraphics[width=8cm,height=5cm]{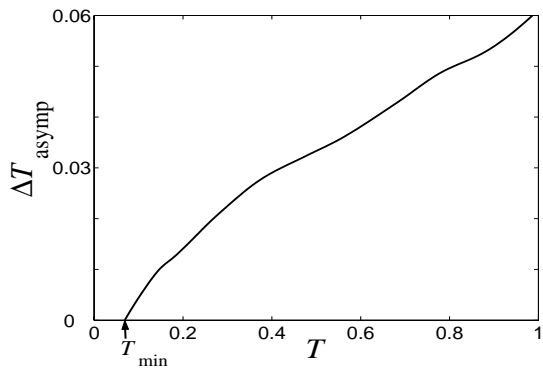}
\caption{ \label{fig5} $\Delta T_{\rm asymp}$ vs. Temperature, for $\Delta E = -0.8$. Refrigeration not possible for $T < T_{\rm min} = 0.065$.}
\end{center}
\end{figure}

\begin{figure}
\begin{center}
\includegraphics[width=8cm,height=5cm]{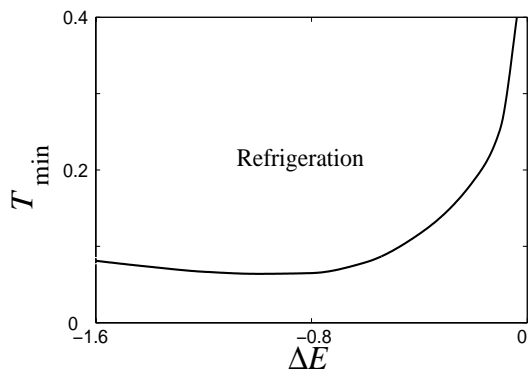}
\caption{ \label{fig6} Minimum operating temperature as a function of the
di.erence in on-site energy }
\end{center}
\end{figure}

The following are some realistic values for the parameters
that we have used here. For a molecular system,
$w = 0.1$ eV, and since we chose $h\nu = w$,
this gives a laser frequency near 20 THz, which is in
the mid-infrared region. Using this and Boltzmann's
constant, we get that our basic unit of heat power
is $[2h\nu^2 ] = 6.4 \times 10^{-7}$ W, and our temperature unit
is $[h\nu/k] = 1160$ K. Room temperature is near
$T = 0.25 ~[h\nu/k]$. With this, the minimum temperature
for operation of this molecular heat pump gives
$T_{min} = 0.065 ~[h\nu/k] = 75 $ K. For a typical distance between
sites in a molecule of $5\rm{\AA}$,  a local electric field amplitude
of $2 \times 10^6 $ V/cm gives $A = 1~[h\nu] = 0.1$ eV. These are values within the range used in experiments of
molecular conduction. In an alternative set up using quantum dots, the corresponding numbers are $w = h\nu = 30~ \mu $eV, which gives a maser frequency in the upper microwave range; the heat flux unit gives $[2h\nu^2 ] = 3\times 10^{-13} $ W, and the unit of temperature gives $[h\nu/k] = 0.35 $ K. With this,
the minimum operating temperature for a quantum-dot based pump would be $T_{\rm min} = 23$ mK.

In this work we have shown that a laser field can produce important effects on the heat transport properties of conducting nanoscale devices. More specifically, we propose that some molecular transistors (ballistic, asymmetrical electron potential) can pump heat when illuminated with a laser field.
As opposed to adiabatic heat pumping, our method does not require any time-dependent modification of the
system's parameters.
There are limitations in the heat pumping capacity of our device. As we have shown in this work, the amplitude
of the laser field needed for the onset of refrigeration grows with the temperature
difference between reservoirs. Also for temperatures below room temperature, there seems to be a maximum temperature difference beyond which we found no refrigeration for any field amplitude.
In addition, there is an absolute minimum value of the operating temperature of the device below which no
heat-pumping action was found to occur. This minimum operating temperature in our two-level device is dependent
on the on-site energy difference ($\Delta E$).

We have only considered here the effect of a laser field on the
electronic contribution to heat transport in a conducting ballistic device. In real systems, the laser field might also produce a perturbation on the phonon contribution (or the vibrational levels contribution) to heat transport. When this contribution is small we can expect these devices to work also as heat pumps, possibly with a more restricted set of parameters.

D.F. Martinez would like to thank Fernando Sols for pointing out Ref. 1.
This work was supported in part from grants from the Hong Kong
Research Grants Council and Hong Kong Baptist University.


\end{document}